\documentclass[aps,prl,superscriptaddress,showpacs,floatfix,twocolumn]{revtex4}

\usepackage{graphicx}   

\def\sqrts{$\sqrt{s_{NN}}$ }
\def\pt{$p_{\rm T}$ }
\def\gevc{GeV/$c$ }
\def\raa{$R_{\rm AA}$ }
\def\auau{Au+Au }
\def\pp{$p$+$p$ }

\begin{document}

\title{Energy Loss and Flow of Heavy Quarks in \auau Collisions at 
       \sqrts = 200 GeV}

\newcommand{\abilene}{Abilene Christian University, Abilene, TX 79699, U.S.}
\newcommand{\banaras}{Department of Physics, Banaras Hindu University, Varanasi 221005, India}
\newcommand{\bnl}{Brookhaven National Laboratory, Upton, NY 11973-5000, U.S.}
\newcommand{\caucr}{University of California - Riverside, Riverside, CA 92521, U.S.}
\newcommand{\charlesczech}{Charles University, Ovocn\'{y} trh 5, Praha 1, 116 36, Prague, Czech Republic}
\newcommand{\ciae}{China Institute of Atomic Energy (CIAE), Beijing, People's Republic of China}
\newcommand{\cns}{Center for Nuclear Study, Graduate School of Science, University of Tokyo, 7-3-1 Hongo, Bunkyo, Tokyo 113-0033, Japan}
\newcommand{\colorado}{University of Colorado, Boulder, CO 80309, U.S.}
\newcommand{\columbia}{Columbia University, New York, NY 10027 and Nevis Laboratories, Irvington, NY 10533, U.S.}
\newcommand{\czechtech}{Czech Technical University, Zikova 4, 166 36 Prague 6, Czech Republic}
\newcommand{\dapnia}{Dapnia, CEA Saclay, F-91191, Gif-sur-Yvette, France}
\newcommand{\debrecen}{Debrecen University, H-4010 Debrecen, Egyetem t{\'e}r 1, Hungary}
\newcommand{\elte}{ELTE, E{\"o}tv{\"o}s Lor{\'a}nd University, H - 1117 Budapest, P{\'a}zm{\'a}ny P. s. 1/A, Hungary}
\newcommand{\fit}{Florida Institute of Technology, Melbourne, FL 32901, U.S.}
\newcommand{\fsu}{Florida State University, Tallahassee, FL 32306, U.S.}
\newcommand{\gsu}{Georgia State University, Atlanta, GA 30303, U.S.}
\newcommand{\hiroshima}{Hiroshima University, Kagamiyama, Higashi-Hiroshima 739-8526, Japan}
\newcommand{\ihepprot}{IHEP Protvino, State Research Center of Russian Federation, Institute for High Energy Physics, Protvino, 142281, Russia}
\newcommand{\illuiuc}{University of Illinois at Urbana-Champaign, Urbana, IL 61801, U.S.}
\newcommand{\instpasczech}{Institute of Physics, Academy of Sciences of the Czech Republic, Na Slovance 2, 182 21 Prague 8, Czech Republic}
\newcommand{\isu}{Iowa State University, Ames, IA 50011, U.S.}
\newcommand{\jinrdubna}{Joint Institute for Nuclear Research, 141980 Dubna, Moscow Region, Russia}
\newcommand{\kaeri}{KAERI, Cyclotron Application Laboratory, Seoul, South Korea}
\newcommand{\kek}{KEK, High Energy Accelerator Research Organization, Tsukuba, Ibaraki 305-0801, Japan}
\newcommand{\kfki}{KFKI Research Institute for Particle and Nuclear Physics of the Hungarian Academy of Sciences (MTA KFKI RMKI), H-1525 Budapest 114, POBox 49, Budapest, Hungary}
\newcommand{\korea}{Korea University, Seoul, 136-701, Korea}
\newcommand{\kurchatov}{Russian Research Center ``Kurchatov Institute", Moscow, Russia}
\newcommand{\kyoto}{Kyoto University, Kyoto 606-8502, Japan}
\newcommand{\labllr}{Laboratoire Leprince-Ringuet, Ecole Polytechnique, CNRS-IN2P3, Route de Saclay, F-91128, Palaiseau, France}
\newcommand{\lawllnl}{Lawrence Livermore National Laboratory, Livermore, CA 94550, U.S.}
\newcommand{\losalamos}{Los Alamos National Laboratory, Los Alamos, NM 87545, U.S.}
\newcommand{\lpc}{LPC, Universit{\'e} Blaise Pascal, CNRS-IN2P3, Clermont-Fd, 63177 Aubiere Cedex, France}
\newcommand{\lund}{Department of Physics, Lund University, Box 118, SE-221 00 Lund, Sweden}
\newcommand{\muenster}{Institut f\"ur Kernphysik, University of Muenster, D-48149 Muenster, Germany}
\newcommand{\myongji}{Myongji University, Yongin, Kyonggido 449-728, Korea}
\newcommand{\nagasaki}{Nagasaki Institute of Applied Science, Nagasaki-shi, Nagasaki 851-0193, Japan}
\newcommand{\newmex}{University of New Mexico, Albuquerque, NM 87131, U.S. }
\newcommand{\nmsu}{New Mexico State University, Las Cruces, NM 88003, U.S.}
\newcommand{\ornl}{Oak Ridge National Laboratory, Oak Ridge, TN 37831, U.S.}
\newcommand{\orsay}{IPN-Orsay, Universite Paris Sud, CNRS-IN2P3, BP1, F-91406, Orsay, France}
\newcommand{\peking}{Peking University, Beijing, People's Republic of China}
\newcommand{\pnpi}{PNPI, Petersburg Nuclear Physics Institute, Gatchina, Leningrad region, 188300, Russia}
\newcommand{\riken}{RIKEN, The Institute of Physical and Chemical Research, Wako, Saitama 351-0198, Japan}
\newcommand{\rikjrbrc}{RIKEN BNL Research Center, Brookhaven National Laboratory, Upton, NY 11973-5000, U.S.}
\newcommand{\rikkyo}{Physics Department, Rikkyo University, 3-34-1 Nishi-Ikebukuro, Toshima, Tokyo 171-8501, Japan}
\newcommand{\saispbstu}{Saint Petersburg State Polytechnic University, St. Petersburg, Russia}
\newcommand{\saopaulo}{Universidade de S{\~a}o Paulo, Instituto de F\'{\i}sica, Caixa Postal 66318, S{\~a}o Paulo CEP05315-970, Brazil}
\newcommand{\seoulnat}{System Electronics Laboratory, Seoul National University, Seoul, South Korea}
\newcommand{\stonybrkc}{Chemistry Department, Stony Brook University, Stony Brook, SUNY, NY 11794-3400, U.S.}
\newcommand{\stonycrkp}{Department of Physics and Astronomy, Stony Brook University, SUNY, Stony Brook, NY 11794, U.S.}
\newcommand{\subatech}{SUBATECH (Ecole des Mines de Nantes, CNRS-IN2P3, Universit{\'e} de Nantes) BP 20722 - 44307, Nantes, France}
\newcommand{\tenn}{University of Tennessee, Knoxville, TN 37996, U.S.}
\newcommand{\titech}{Department of Physics, Tokyo Institute of Technology, Oh-okayama, Meguro, Tokyo 152-8551, Japan}
\newcommand{\tsukuba}{Institute of Physics, University of Tsukuba, Tsukuba, Ibaraki 305, Japan}
\newcommand{\vandy}{Vanderbilt University, Nashville, TN 37235, U.S.}
\newcommand{\waseda}{Waseda University, Advanced Research Institute for Science and Engineering, 17 Kikui-cho, Shinjuku-ku, Tokyo 162-0044, Japan}
\newcommand{\weizmann}{Weizmann Institute, Rehovot 76100, Israel}
\newcommand{\yonsei}{Yonsei University, IPAP, Seoul 120-749, Korea}
\affiliation{\abilene}
\affiliation{\banaras}
\affiliation{\bnl}
\affiliation{\caucr}
\affiliation{\charlesczech}
\affiliation{\ciae}
\affiliation{\cns}
\affiliation{\colorado}
\affiliation{\columbia}
\affiliation{\czechtech}
\affiliation{\dapnia}
\affiliation{\debrecen}
\affiliation{\elte}
\affiliation{\fit}
\affiliation{\fsu}
\affiliation{\gsu}
\affiliation{\hiroshima}
\affiliation{\ihepprot}
\affiliation{\illuiuc}
\affiliation{\instpasczech}
\affiliation{\isu}
\affiliation{\jinrdubna}
\affiliation{\kaeri}
\affiliation{\kek}
\affiliation{\kfki}
\affiliation{\korea}
\affiliation{\kurchatov}
\affiliation{\kyoto}
\affiliation{\labllr}
\affiliation{\lawllnl}
\affiliation{\losalamos}
\affiliation{\lpc}
\affiliation{\lund}
\affiliation{\muenster}
\affiliation{\myongji}
\affiliation{\nagasaki}
\affiliation{\newmex}
\affiliation{\nmsu}
\affiliation{\ornl}
\affiliation{\orsay}
\affiliation{\peking}
\affiliation{\pnpi}
\affiliation{\riken}
\affiliation{\rikjrbrc}
\affiliation{\rikkyo}
\affiliation{\saispbstu}
\affiliation{\saopaulo}
\affiliation{\seoulnat}
\affiliation{\stonybrkc}
\affiliation{\stonycrkp}
\affiliation{\subatech}
\affiliation{\tenn}
\affiliation{\titech}
\affiliation{\tsukuba}
\affiliation{\vandy}
\affiliation{\waseda}
\affiliation{\weizmann}
\affiliation{\yonsei}
\author{A.~Adare}	\affiliation{\colorado}
\author{S.~Afanasiev}	\affiliation{\jinrdubna}
\author{C.~Aidala}	\affiliation{\columbia}
\author{N.N.~Ajitanand}	\affiliation{\stonybrkc}
\author{Y.~Akiba}	\affiliation{\riken} \affiliation{\rikjrbrc}
\author{H.~Al-Bataineh}	\affiliation{\nmsu}
\author{J.~Alexander}	\affiliation{\stonybrkc}
\author{A.~Al-Jamel}	\affiliation{\nmsu}
\author{K.~Aoki}	\affiliation{\kyoto} \affiliation{\riken}
\author{L.~Aphecetche}	\affiliation{\subatech}
% ...
\collaboration{PHENIX Collaboration} \noaffiliation

\date{\today}

\begin{abstract}
The PHENIX experiment at the Relativistic Heavy Ion Collider (RHIC)
has measured electrons with $0.3 < p_{\rm T} < 9$ GeV/$c$ at midrapidity 
($|y| < 0.35$) from heavy flavor (charm and bottom) decays in \auau
collisions at \sqrts = 200 GeV.
The nuclear modification factor $R_{\rm AA}$ relative to \pp collisions
shows a strong suppression in central \auau collisions, indicating 
substantial energy loss of heavy quarks in the medium produced at RHIC
energies.
A large azimuthal anisotropy, $v_2$, with respect to the reaction plane is
observed for $0.5 < p_{\rm T} < 5$ GeV/$c$ indicating substantial heavy 
flavor elliptic flow.
Both \raa and $v_2$ show a \pt dependence different from those of neutral
pions.  
A comparison to transport models which simultaneously describe 
$R_{\rm AA}(p_{\rm T})$ and $v_2(p_{\rm T})$ suggests that the viscosity 
to entropy density ratio is close to the conjectured quantum lower bound, 
{\it i.e.} near a perfect fluid.
\end{abstract}


\maketitle

Experimental results from the Relativistic Heavy Ion Collider (RHIC)
have established that dense partonic matter is formed in Au+Au
collisions at RHIC~\cite{wp_phenix,wp_brahms,wp_phobos,wp_star}.
Strong suppression observed for $\pi^0$ and other light hadrons at
high transverse momentum ($p_{\rm T}$)~\cite{ppg003,ppg014,sup_star,ppg051}
indicates partonic energy loss in the produced medium.
The azimuthal anisotropy $v_2(p_{\rm T})$~\cite{ppg022,star_v2_1} provides 
evidence that collective motion develops in a very early stage of the 
collision ($\tau\lesssim$ 5 fm/$c$), in accordance with hydrodynamical 
calculations~\cite{hydro,Hirano:2005wx}.
The comparison of $v_2$ with several such models
suggests~\cite{Shuryak:2003xe,Gyulassy:2004zy,Kolb:2003dz} that the matter 
formed at RHIC is a near-perfect fluid with viscosity to entropy density ratio 
$\eta/s$ close to the conjectured quantum lower bound~\cite{Kovtun:2004de}.
Energy loss and flow are related to the transport properties of the medium
at temperature $T$, in particular the diffusion coefficient 
$D \propto \eta/(sT)$.

Further insight into properties of the medium can be gained from the 
production and propagation of particles carrying heavy quarks 
(charm or bottom).
A fixed-order-plus-next-to-leading-log (FONLL) 
perturbative QCD (pQCD)
calculation~\cite{fonll} describes the cross sections of
heavy-flavor decay electrons in \pp collisons at $\sqrt{s} = 200$~GeV 
within theoretical uncertainties~\cite{ppg065}.
In \auau collisions the total yield of such electrons was found to scale 
with the number of nucleon-nucleon collisions as expected for point-like 
processes~\cite{ppg035}. 
Energy loss via gluon radiation is expected to be reduced for heavy quarks 
due to suppression of forward radiation, thus increasing their expected 
thermalization time~\cite{dk,djord,wied}.
Consequently, a decrease of high \pt suppression and of $v_2$ is 
expected from light to charm to bottom quarks, with the absolute values 
and their \pt dependence sensitive to the properties of the medium. 
In contrast to these expectations a strong suppression of heavy-flavor 
decay electrons was discovered for $2 < p_{\rm T} < 5$ 
GeV/$c$~\cite{ppg056,star_e}, together with nonzero electron $v_2$ for 
$p_{\rm T} < 2$ GeV/$c$~\cite{ppg040}.

This Letter presents \pt spectra and the elliptic flow amplitude 
$v_2^{\rm HF}$ of electrons, $(e^++e^-)/2$, from heavy-flavor decays 
at midrapidity in \auau collisions at \sqrts = 200 GeV.
An increase in statistics by more than a factor ten and reduced systematic 
uncertainties compared to earlier data~\cite{ppg035,ppg040,ppg056} greatly
extend the $p_{\rm T}$ range both for the determination of the centrality
dependence of $R_{\rm AA}$ and for the measurement of $v_2^{\rm HF}$.

The data were collected by the PHENIX detector~\cite{phenix} in the 2004
RHIC run. The minimum bias trigger and the collision centrality
were obtained from the beam-beam counters (BBC) and zero degree
calorimeters~\cite{wp_phenix}.
After selecting good runs, data samples of 8.1 and 7.0 $\times$ 10$^8$ 
minimum bias events in the vertex range $|z_{\rm vtx}| < 20$~cm are used for 
the spectra and $v_2$ analyses, respectively.

Charged particle tracks are reconstructed with the two PHENIX central arm
spectrometers, each covering $\Delta\phi = \pi/2$ in azimuth and
$|\eta| < 0.35$ in pseudo-rapidity~\cite{phenix}.
Tracks are confirmed by matching showers in the electromagnetic calorimeter
(EMCal) within $2\sigma$ in position.
Electron candidates have at least three associated hits in the ring imaging
\v{C}erenkov detectors (RICH) and fulfill a shower shape cut in the EMCal, 
where they deposit an energy, $E$, consistent with the momentum
($E/p - 1 > -2\sigma$).
Below the \v{C}erenkov threshold for pions ($p_{\rm T} < 5$~GeV/$c$)
electron mis-identification is only due to random coincidences between
hadron tracks and hits in the RICH.
This small background ($<20$\% at low \pt in central collisions, less towards
high \pt and peripheral events) is subtracted statistically using an event 
mixing technique.
Requiring at least five hits in the RICH and tightening the shower
shape cut extends the electron measurement to 9~\gevc in $p_{\rm T}$, with
negligible hadron background for $p_{\rm T} < 8$~GeV/$c$ and a hadron
contamination of 20\% for $8 < p_{\rm T} < 9$~GeV/$c$. 
The raw spectra are corrected for geometrical acceptance and reconstruction
efficiency determined by a GEANT simulation. The
centrality dependent efficiency loss $<2$\% ($\approx23$\%) for
peripheral (central) events is evaluated by reconstructing simulated
electrons embedded into real events.

The inclusive electron spectra consist of
(1) ``non-photonic'' electrons from heavy-flavor decays,
(2) ``photonic" background from Dalitz decays and photon conversions 
   (mainly in the beam pipe), and
(3) ``non-photonic'' background from $K\rightarrow e\pi\nu$ ($K_{e3}$) and 
    dielectron decays of vector mesons.
Contribution (3) is small ($<$10\% for $p_{\rm T}<$ 0.5~GeV/$c$, $<$2\% for 
$p_{\rm T}>$ 2~GeV/$c$) compared to (2).
The heavy-flavor signal and the ratio of non-photonic to photonic 
electrons, $R_{\rm NP}$, are determined via two independent and complementary 
methods described in detail in \cite{ppg065}, where the identical detector 
configuration was used. 
At low $p_{\rm T}$ ($p_{\rm T} < 1.6$ GeV/$c$), where the heavy-flavor 
signal to background ratio is small (S/B $<$ 1), the ``converter subtraction'' 
method is used which employs a photon converter of 1.67\% radiation length 
($X_0$) installed around the beam pipe for part of the run. 
The converter multiplies the photonic background by a known, nearly \pt 
independent factor $R_\gamma \sim 2.3$.
The photonic background can then be determined by comparing the inclusive
electron yield with and without the converter.
For higher $p_{\rm T}$, where S/B is large, the ``cocktail subtraction" 
method~\cite{ppg056} is used. 
Here the background is calculated with a Monte Carlo hadron
decay generator and subtracted from the data. 
At low \pt the dominant background source is the $\pi^0$ Dalitz decay, 
which is calculated for each centrality using measured pion 
spectra~\cite{ppg014,ppg026} as input.
In good agreement with measured data~\cite{ppg051}, the spectral shapes of 
other light hadrons $h$ ($\eta$, $\rho$, $\omega$, $\phi$, $\eta'$) are 
derived from the pion spectrum assuming a universal shape in 
$m_T = \sqrt{p_{\rm T}^2 +m_h^2}$ with a fixed constant ratio at high 
$p_{\rm T}$.
Photon conversions in the beam pipe, air and helium bags 
(total: $0.4\% X_0$) are also included, along with background from $K_{e3}$ 
decays and both external and internal conversions of direct photons
which are important for $p_{\rm T} > 4$~GeV/$c$.
The agreement within the systematic uncertainties in the overlap region 
$0.3 < p_{\rm T} < 4$~GeV/$c$ of these two methods demonstrates that the 
absolute value of photonic backgrounds in the PHENIX aperture is 
well-understood.

The $v_2$ of inclusive electrons, $v_2^{inc}$, is measured as
$v_2^{inc}=\langle \cos[2(\phi-\Phi_R)] \rangle / \sigma_R$~\cite{rpm},
where $\Phi_R$ is the azimuthal orientation of the reaction plane
measured with the resolution $\sigma_R$ using the BBC~\cite{ppg022}.
Since $\sigma_R$ is centrality dependent, $v_2$ is determined for narrow
centrality bins (10\%) and then averaged to calculate $v_2$
for minimum bias events.
The $v_2$ of random hadronic background is subtracted statistically as 
described in~\cite{ppg040}.

The $v_2^{non-\gamma}$ of non-photonic electrons is obtained by
subtracting the photonic electron $v_2^\gamma$ as:
$v_2^{non-\gamma} = ((1+R_{\rm NP})v_2^{inc}-v_2^\gamma)/R_{NP}$.
Here $v_{2}^\gamma$ is calculated via a Monte Carlo generator that
includes $\pi^0$, $\eta$, and direct photons.
The measured $v_2(p_{\rm T})$ of $\pi^\pm$,$\pi^0$ and 
$K^\pm$~\cite{ppg022,ppg046}
is used as input, assuming $v_2^{\pi^\pm}=v_2^{\pi^0}$,
$v_2^\eta = v_2^{K^\pm}$, and $v_2^{{\rm direct}\gamma}=0$.
A direct measurement of $v_2^\gamma$ using the converter subtraction
method confirms the calculation within statistical uncertainties.
The resulting $v_2^{non-\gamma}$ has a small contribution from $K_{e3}$ 
background which is simulated and subtracted to obtain $v_2^{\rm HF}$ of 
heavy-flavor decay electrons.

Three independent categories of systematic uncertainties are considered.
(A) The inclusive electron spectra include uncertainties in the geometrical 
acceptance (5\%), the reconstruction efficiency (3\%), and the embedding 
correction ($\le$4\%).
(B) Uncertainties in the converter subtraction are mainly given by the
uncertainty in $R_\gamma$ (2.7\%) and in the relative acceptance of runs with
and without the converter being installed (1\%).
(C) Uncertainties in the cocktail subtraction rise from 8\% at
$p_{\rm T} = 0.3$~GeV/$c$ to 13\% at 9~GeV/$c$, dominated by systematic 
errors in the pion input and, at high $p_{\rm T}$, the direct photon spectrum.
The $v_2$ measurement includes a systematic uncertainty of 5\% due to the 
reaction plane uncertainty.

\begin{figure}[tb]
\includegraphics[width=1.0\linewidth]{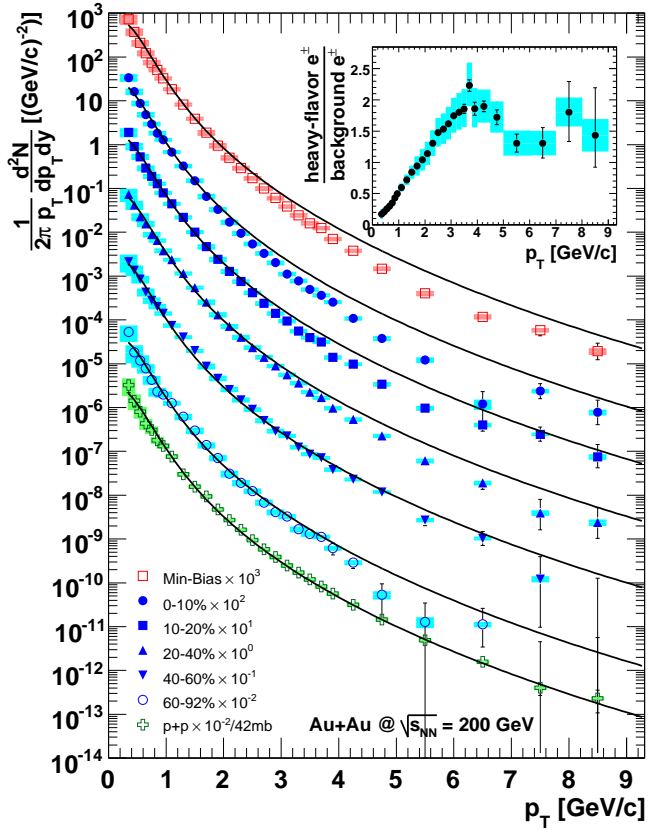}
\caption {\label{fig1}Invariant yields of electrons from heavy-flavor 
decays for different \auau centrality classes and for \pp collisions, 
scaled by powers of ten for clarity. The solid lines are the result 
of a FONLL calculation normalized to the \pp data~\cite{ppg065} and 
scaled with $\langle T_{\rm AA} \rangle$ for each \auau centrality class. 
The insert shows the ratio of heavy-flavor to background electrons 
for minimum bias \auau collisions. Error bars (boxes) depict statistical 
(systematic) uncertainties.}
\end{figure}

Figure~\ref{fig1} shows the invariant \pt spectra of electrons from 
heavy-flavor decay for minimum bias events and in five centrality classes.
The curves overlayed are the fit to the corresponding data from \pp 
collisions~\cite{ppg065} with the spectral shape taken from a FONLL 
calculation~\cite{fonll} and scaled by the nuclear overlap integral 
$\langle T_{\rm AA} \rangle$ for each centrality class~\cite{ppg014}.
The insert in Fig.~\ref{fig1} shows the ratio of electrons from
heavy-flavor decays to background.
It increases rapidly with $p_{\rm T}$, exceeding unity for $p_{\rm T} > 
1.8$~GeV/$c$, reflecting the small amount of material in the detector 
acceptance which makes the accurate measurement of heavy-flavor electron 
spectra and $v_2^{\rm HF}$ possible.

For all centralities, the \auau spectra agree well with the \pp reference 
at low \pt but a suppression with respect to \pp develops towards high 
$p_{\rm T}$.
This is quantified by the nuclear modification factor
$R_{\rm AA} = dN_{Au+Au}/(\langle T_{AA} \rangle d\sigma_{p+p})$,
where $dN_{Au+Au}$ is the differential yield in \auau and $d\sigma_{p+p}$
is the differential cross section in \pp in a given \pt bin.
For $p_{\rm T} < 1.6$~GeV/$c$, $d\sigma_{p+p}$, is taken bin-by-bin 
from~\cite{ppg065}, whereas a fit to the same data (curves in Fig.~\ref{fig1})
is used at higher $p_{\rm T}$, taking systematic uncertainties in 
$d\sigma_{p+p}$ and $T_{\rm AA}$ into account.

\begin{figure} [hbt]
\includegraphics[width=1.0\linewidth]{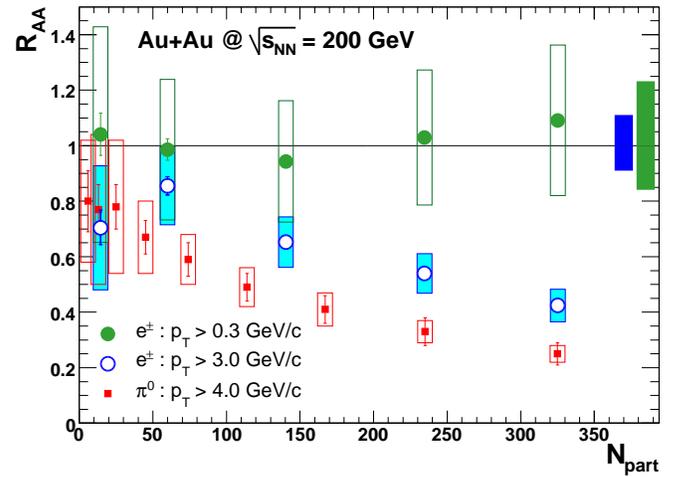}
\caption {\label{fig2}$R_{\rm AA}$ of heavy-flavor electrons with \pt above 
0.3 and 3~GeV/$c$ and of $\pi^0$ with $p_{\rm T} > 4$~GeV/$c$ as function 
of centrality given by $N_{\rm part}$. Error bars (boxes) depict statistical 
(point-by-point systematic) uncertainties. The right (left) box at 
$R_{\rm AA} = 1$ shows the relative uncertainty from the \pp reference 
common to all points for $p_{\rm T} > 0.3 (3)$~GeV/$c$.}
\end{figure}

Figure~\ref{fig2} shows \raa for electrons from heavy-flavor decays
for two different \pt ranges as a function of the number of
participant nucleons, $N_{\rm part}$. 
For the integration interval $p_{\rm T} > 0.3$~GeV/$c$ containing more than 
half of the heavy-flavor decay electrons~\cite{ppg065} \raa is consistent
with unity for all $N_{\rm part}$ in accordance with the binary scaling of 
the total heavy-flavor yield~\cite{ppg035}.
For $p_{\rm T} > 3$~GeV/$c$, the heavy flavor electron \raa decreases 
systematically with centrality, while larger than \raa of $\pi^0$ with 
$p_{\rm T} > 4$~GeV/$c$~\cite{ppg014}.
Since above 3~GeV/$c$ electrons from charm decays originate mainly from $D$ 
mesons with \pt above 4~GeV/$c$ this comparison indicates a smaller
suppression of heavy-flavor mesons than observed for light mesons
in this intermediate $p_{\rm T}$ range.

\begin{figure}[bht]
\includegraphics[width=1.0\linewidth]{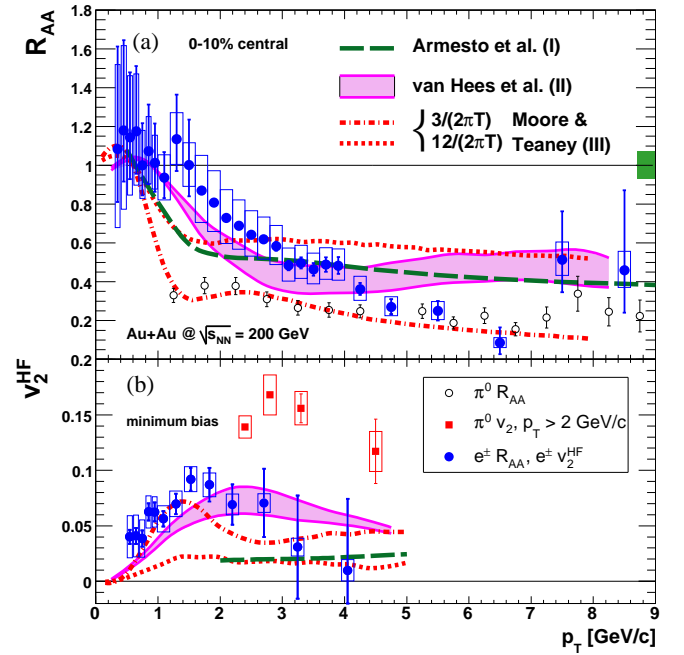}
\caption {\label{fig3}(a) $R_{\rm AA}$ of heavy-flavor electrons 
in 0-10\% central collisions compared with $\pi^0$ data~\cite{ppg014} and 
model calculations (curves I~\cite{Armesto:2005mz}, 
II~\cite{vanHees}, and III~\cite{Moore:2004tg}). 
The box at $R_{\rm AA} = 1$ shows the uncertainty in $T_{AA}$.
(b) $v_2^{\rm HF}$ of heavy-flavor electrons in minimum bias collisions 
compared with $\pi^0$ data~\cite{ppg046} and the same models. 
Errors are shown as in Fig.~\ref{fig2}.}
\end{figure}

Figure~\ref{fig3} shows the measured \raa and $v_2^{\rm HF}$ of 
heavy-flavor electrons in 0-10\% central and minimum bias collisions, 
and our corresponding $\pi^0$ data~\cite{ppg014,ppg046}.
The data indicate strong coupling of heavy quarks to the medium.
While at low $p_{\rm T}$ the suppression is smaller than that of $\pi^0$,
$R_{AA}$ of heavy-flavor decay electrons approaches the $\pi^0$ value for 
$p_{\rm T} > 4$~GeV/$c$ although a significant contribution from bottom decays 
is expected at high $p_{\rm T}$.
The large $v_2^{\rm HF}$ indicates that the charm relaxation time is 
comparable to the short time scale of flow development in the produced medium.
It should be noted that much reduced uncertainties and the extended $p_{\rm T}$
range of the present data permit the comparisons of $R_{AA}$ and $v_2$ of the
heavy and light flavors.

More quantitative statements require theoretical guidance.
Figure~\ref{fig3} compares the $R_{\rm AA}$ and $v_2$ of heavy-flavor 
electrons with models calculating both quantities simultaneously.
A perturbative QCD calculation with radiative energy 
loss (curves I)~\cite{Armesto:2005mz} describes the measured $R_{\rm AA}$ 
reasonably well using a large transport coefficient $\hat{q} = 14$~GeV$^2$/fm, 
which also provides a consistent description of light hadron suppression.
This value of $\hat{q}$ would imply a strongly coupled medium.   
In this model the azimuthal anisotropy is only due to the path length 
dependence of energy loss, and the data clearly favor larger $v_2^{\rm HF}$ 
than predicted from this effect alone.

Figure ~\ref{fig3} also shows that the large $v_2^{\rm HF}$ is better 
reproduced in Langevin-based 
heavy quark transport calculations~\cite{vanHees,Moore:2004tg}.
A calculation which includes elastic scattering mediated by resonance 
excitation (curves II)~\cite{vanHees} is in good agreement with both the 
measured \raa and $v_2$.
This is achieved with a small heavy quark relaxation time $\tau$ which 
translates into a diffusion coefficient $D_{HQ} \times (2\pi T) = 4$-$6$
in this model~\cite{vanHees}.
Energy loss and flow are also calculated in~\cite{Moore:2004tg} in terms 
of $D_{HQ}$ (curves III).
While this model fails to simultaneously describe the measured \raa and $v_2$ 
with one value for $D_{HQ}$, the range for $D_{HQ}$ leading to reasonable 
agreement with \raa or $v_2$ is similar to that from~\cite{vanHees}, again
implying that small $\tau$ and/or $D_{HQ} \times (2\pi T)$ are required 
to reproduce the data.
Note that $D_{HQ}$ provides an upper bound for the bulk matter's diffusion 
coefficient $D$.
Using the observation~\cite{Moore:2004tg} that
$D \approx 6 \times \eta/(\epsilon+p)$ with $\epsilon+p = Ts$ at $\mu_B=0$
provides an estimate for the viscosity to entropy ratio
$\eta/s \approx (\frac{4}{3}-2)/4\pi$, intriguingly close to the conjectured 
quantum lower bound $1/4\pi$~\cite{bound}.
This result is consistent with estimates obtained in the light quark
sector from elliptic flow~\cite{roy} and fluctuation analyses~\cite{gavin}.

The conjecture of a bound on $\eta/s$~\cite{Kovtun:2004de} was obtained 
using the anti-de Sitter-space/conformal-field-theory
correspondence~\cite{Maldacena:1997re,Witten:1998zw},
which exploits a duality between strongly coupled gauge theories and
semiclassical gravitational physics.
Recently, such methods were applied to estimate
$\hat{q}$\cite{Liu} and $D_{HQ}$ in a 
thermalized plasma~\cite{Herzog:2006gh,Gubser:2006bz,Friess:2006aw}.
These authors also find a small diffusion coefficient 
$D_{HQ} \times (2\pi T) \sim 1$.

In conclusion, we have observed large energy loss and flow of heavy quarks 
in \auau collisions at \sqrts = 200 GeV.
The data provide strong evidence for the coupling of heavy quarks to the 
produced medium.
A short relaxation time of heavy quarks and/or a small diffusion coefficient
are required by the data. 
A model comparison suggests a viscosity to entropy ratio of the medium 
close to the quantum lower bound, {\it i.e.} near a perfect fluid.


We thank the staff of the Collider-Accelerator and 
Physics Departments at BNL for their vital contributions.  
We acknowledge support from 
the Department of Energy and NSF (U.S.A.), 
MEXT and JSPS (Japan), 
CNPq and FAPESP (Brazil), 
NSFC (China), 
MSMT (Czech Republic),
IN2P3/CNRS, and CEA (France), 
BMBF, DAAD, and AvH (Germany), 
OTKA (Hungary), 
DAE (India), 
ISF (Israel), 
KRF and KOSEF (Korea), 
MES, RAS, and FAAE (Russia),
VR and KAW (Sweden), 
U.S. CRDF for the FSU, 
US-Hungarian NSF-OTKA-MTA, 
and US-Israel BSF.

\def\IJMPA{{Int. J. Mod. Phys.}~{\bf A}}
\def\JPG{{J. Phys}~{\bf G}}
\def\NCA{Nuovo Cimento}
\def\NIM{Nucl. Instrum. Methods}
\def\NIMA{{Nucl. Instrum. Methods}~{\bf A}}
\def\NPA{{Nucl. Phys.}~{\bf A}}
\def\NPB{{Nucl. Phys.}~{\bf B}}
\def\PLB{{Phys. Lett.}~{\bf B}}
\def\PLC{Phys. Repts.\ }
\def\PRL{Phys. Rev. Lett.\ }
\def\PRC{{Phys. Rev.}~{\bf C}}
\def\PRD{{Phys. Rev.}~{\bf D}}
\def\ZPC{{Z. Phys.}~{\bf C}}
\def\etal{{\it et al.}}

\end{document}